\documentclass[prd,a4paper,aps,epsfig,showpacs,twocolumn,floats]{revtex4}
\usepackage{epsfig}

\newcommand{\be}{\begin{equation}}
\newcommand{\ee}{\end{equation}}
\newcommand{\bea}{\begin{eqnarray}}
\newcommand{\eea}{\end{eqnarray}}

\begin{document}
\title{Observing the temperature of the Big Bang through large scale structure}
\author{Pedro G. Ferreira$^1$ and Jo\~ao  Magueijo$^{2,3,4}$}
\affiliation{$^1$Astrophysics, University of Oxford,
Oxford, OX1 3RH\\
$^2$Perimeter Institute for Theoretical Physics, 31
Caroline St N, Waterloo N2L 2Y5, Canada\\
$^3$ Canadian Institute for Theoretical Astrophysics,
60 St George St, Toronto M5S 3H8, Canada\\
$^4$ Theoretical Physics, Imperial College, London, SW7 2BZ}
\date{\today}
\begin{abstract}
{It is widely accepted that the Universe underwent a period of
thermal equilibrium at very early times. One expects a residue of
this primordial state to be imprinted on the large scale structure
of space time. In this  paper we study the morphology of this
thermal residue in a universe whose early dynamics is governed by
a scalar field. We calculate the amplitude of fluctuations on
large scales and compare it to the imprint of vacuum fluctuations.
We then use the observed power spectrum of fluctuations on the
cosmic microwave background to place a constraint on the
temperature of the Universe before and during inflation. We also
present an alternative scenario where the fluctuations are
predominantly thermal and near scale-invariant.}
\end{abstract}
\pacs{0000000}
\maketitle

\bigskip

A cornerstone of modern cosmology is that the universe underwent a
sustained period of thermal equilibrium at early times. Two of the
key predictions of the big bang cosmology, the spectrum of the
cosmic microwave background and the abundance of light elements,
hinge on the existence of this primordial hot phase \cite{Peacock}. A key
characteristic of systems in thermal equilibrium is the presence
of fluctuations. These are, to some extent, uniquely defined and
can be derived from the microphysical properties of the system \cite{LanLif}.
Hence we expect to be able to characterize the fluctuations of the
energy density of the early universe which in turn lead to
irregularities in the fabric in space time. These should be
reflected in the distribution of large scale structure, the
propagation of light rays and other such cosmological observables.

There have been a number of attempts at pinning down the fine
details of the thermal fluctuations in the early universe. Under
standard assumptions it can be shown that thermal models are
observationally unsound. A generic $n_s=4$ prediction for the
spectral index follows, unless there is a phase transition, in
which case $n_s=0$. This can be bypassed, and a more congenial
$n_s\approx 1$ be predicted, by considering non-standard
assumptions: e.g. by considering a gas of strings at the Hagedorn
phase~\cite{hag}, or by invoking an early holographic phase in
loop quantum cosmology~\cite{holo}, followed by a phase
transition. One can also appeal to the technicalities of loop
quantum cosmology~\cite{param} or postulate a mildly sub-extensive
contribution to the energy density~\cite{milne}. All these
scenarios require speculative new physics.

In this paper, we revisit this issue by focusing on what has become
a standard and
fruitful model of the universe: a perturbed homogeneous and isotropic
spacetime whose dynamics is driven by a scalar field. Without loss of
generality, we will restrict
ourselves to a scalar field with an exponential potential but will
allow both positive and negative kinetic energies \cite{exp}.
If the field rolls sufficiently
slowly away from the origin, we have power law, accelerated expansion.
If the field rolls sufficiently quickly, the energy density in the
scalar field will mimic the behaviour of an assortment of cosmological
fluids (such as radiation or dust). If the kinetic energy of the scalar field
is negative, we obtain ``phantom''-like behaviour: the effective equation
of state $w\equiv P/\rho$ (where $\rho$ and $P$ are the energy density and
pressure in the scalar field) is such that $w<-1$. Such a setup
allows us to analytically calculate the amplitude and spectrum of thermal
fluctuations including gravitational backreaction.
In this paper we will
focus on universes that underwent superluminal expansion.

Let us briefly revisit the model. We will consider a potential for
the scalar field of the form:
$V(\phi)=M_{Pl}^4\exp(-\sqrt{\frac{2}{p}}\frac{\phi}{M_{Pl}})$
where $M_{Pl}$ is the reduced Planck mass. The evolution of the
scalar field is given by $\phi=\sqrt{2p}M_{Pl}\ln
({M_{Pl}t}/{\sqrt{p(3p-1)}})$ and the Friedman equations lead to a
simple solution of the form $a\propto t^p$ and $H\equiv{\dot
a}{a}= {p}/{t}$ where $^.\equiv d/dt$. Note that, if $p>1$, the
expansion is superluminal. It is convenient to rewrite some of
these results in terms of conformal time, $\tau$. If $p>1$ we have
that the past is at $\tau=-\infty$ and blows up at $\tau=0$. We
then have that the scale factor and the conformal Hubble parameter
is given by $a\propto(-\tau)^{\frac{-p}{p-1}}$ and the conformal
Hubble parameter is given by ${\cal H}\equiv
\frac{a'}{a}=\frac{-p}{p-1}\frac{1}{\tau}$ where $'=d/d\tau$.

Let us now focus on how perturbations on these background cosmologies are
seeded and evolve \cite{mukh}. Recall
that we can expand a scalar field and space-time metric
around a homogeneous background, $\phi=\phi_0+\varphi$ and
$ds^2=a^2[(1+2\Phi)d\tau^2-(1-2\Psi)d{\bf r}^2]$. The quantity
of choice is the gauge invariant  variable,
\begin{eqnarray}
v=a(\delta\varphi+\frac{{\dot \phi}_0}{H}\Psi) \nonumber
\end{eqnarray}
which can be related to the curvature perturbation, $ {\cal
R}=-{v}/{z}$ where $z=\frac{a{\dot \phi}_0}{H}$ The gauge
invariant Newtonian (or ``Bardeen'') potential, $\Phi$, can be
found from ${\cal R}$ through $k^2\Phi=4\pi G {\dot \phi}_0z{\cal
R}' $ For our choice of background cosmologies, the gauge
invariant perturbation variable obeys a Bessel equation with a
general solution given by:
\begin{eqnarray}
v_{\bf k}(\tau)=A_{\bf k}(|\tau|)^{1/2}J_{\nu}(k|\tau|)+
B_{\bf k}(|\tau|)^{1/2}Y_{\nu}(k|\tau|) \nonumber
\end{eqnarray}
where $J_\nu(x)$ and $Y_\nu(x)$ are Bessel functions with $\nu=\frac{3}{2}+\frac{1}{p-1}$.

In a universe undergoing superluminal expansion there is a natural
mechanism by which fluctuations can be seeded. We assume that
$v$ is promoted to a quantum operator:
\begin{eqnarray}
{\hat v}(\tau,{\bf x})= \int \frac{d^3 {\bf k}}{(2\pi)^{3/2}}
[v_k(\tau){\hat a}_{\bf k}e^{{i\bf k}\cdot {\bf x}}+
v^*_k(\tau){\hat a}^\dagger_{\bf k}e^{{-i\bf k}\cdot {\bf x}}] \nonumber
\end{eqnarray}
where ${\hat a}_{\bf k}$ and its conjugate are the annihilation
and creation operators for the $n$ particle state. We are
interested in the two point correlation function $\langle {\hat
v}({\bf x}+{\bf r}){\hat v}({\bf x})\rangle$, where
$\langle\cdots\rangle$ is a quantum expectation value depends on
the state one is considering, i.e. $\langle A \rangle= \langle
\zeta |{\hat A}| \zeta \rangle$. A natural choice is the
ground-state or vacuum state of each mode, $|\zeta\rangle
=|0\rangle$. In the past, where $(-k\tau)\rightarrow -\infty$, a
given mode was well within the horizon. This allows us to uniquely
define the solution (i.e. the coefficients $A_{\bf k}$ and $B_{\bf
k}$) to the mode equation to be $v_{\bf
k}(\tau)=\frac{\sqrt{\pi}}{2}e^{i(\nu+1/2)\pi/2}(-\tau)^{1/2}H^{(1)}_\nu(-k\tau)$
(where $H^{(1)}_\nu(x)$ is a Hankel function). This solution has a
unique behaviour at late times (i.e. when $(-k\tau)\rightarrow
0$):
\begin{eqnarray}
v_{\bf k}(\tau)\rightarrow e^{i(\nu-1/2)\pi/2}2^{\nu-3/2}&&\frac{\Gamma[\nu]}
{\Gamma[3/2]}\frac{1}{\sqrt{2k}}(-k\tau)^{-\nu+1/2}
\nonumber \\ &&\times [1-\frac{(-k\tau)^2}{4(1-\nu)}]
 \nonumber
\end{eqnarray}
${\cal R}$ can be trivially obtained from the above solution and we find that
\begin{eqnarray}
{\cal P}^0_{\cal R}(k)=\frac{2^{2\nu-4}}{2\pi^2}\left(\frac{\Gamma[\nu]}
{\Gamma[3/2]}\right)^2\frac{(-\tau)^{-2\nu+1}}{z^2}k^{-2/(p-1)} \label{vacspec}
\end{eqnarray}
which goes to a constant as $p\rightarrow \infty$. We find the
well known result that the scalar spectral index is given by
$n_S-1=\frac{2}{1-p}=\frac{6(1+w)}{1+3w}$. We can see
that, in the limit of $w\rightarrow -1$ we have pure scale
invariance.

Throughout the above calculation we have discarded any reference
to the hot origins of the universe. Yet we are starting off at
high energies, when the Universe would have been strongly
interacting. It would be natural to expect the imprint of these
thermal initial conditions on the scalar field in some way. Indeed
one would expect fluctuations in the scalar field to be
thermalized through a variety of different mechanisms. The
universe may have entered a scalar field dominated regime from a
preceding radiation dominated regime; interactions with the hot
radiation would have led the fluctuations in the scalar field to
be thermal. Furthermore, the scalar field model we are considering
has non-linear self-interactions through the exponential
potential. Very short wave modes would play the role of a heat
bath even through the period of superluminal expansion and scalar
field domination. The details of how primordial fields undergo
evolution in a hot phase have been studied in great detail in
\cite{Boy,warm} where a number of effects where identified
emerging from the non-equilibrium nature of the problem.

In what follows, we will disregard non-equilibrium effects: these will
introduce small corrections and can be included in a more detail calculation.
Our calculation is therefore undertaken in the setting of equilibrium
statistical mechanics:
the appropriate expectation value to consider
is given by $\langle A \rangle= \sum_{n} \rho_{nn}\langle n|{\hat A}|n\rangle/
(\sum_{n} \rho_{nn}\langle n|n\rangle)$, where $|n\rangle$ is the
n-particle state (referring to a given momentum ${\bf k}$).
 The simplest approach is to
simply posit that each mode is Boltzman weighted. Recall that this involves
setting the density matrix above to $\rho_{nn}=e^{-\beta E_n}$ where $E_n$ is the
energy of a given mode with occupation number $n$, $\beta=1/K_BT$, $K_B$ is the Boltzman constant and $T$ is the
temperature.
Hence we find that
\begin{eqnarray}
\langle {\hat v}({\bf x}+{\bf r}){\hat v}({\bf x})\rangle=
\int \frac{d^3 {\bf k}}{(2\pi )^{3/2}}|v_k(\tau)|^2[2n(k)+1]e^{i{\bf k}\cdot r}
\nonumber
\end{eqnarray}
where the resulting number density (subtracting out the vacuum
state) for  is given by $n(k)=\frac{1}{e^{\beta E(k,\tau)}-1}$.

The energy of the perturbation can be found from the Hamiltonian
density of $v$. With the above solutions we have that
$E(k,\tau)=\frac{\hbar \pi|\tau|}{4}B(k,\tau)|H_\nu(k|\tau|)$
with $B(k,\tau)=(3p^2-p)/[\tau^2(p-1)^2]+k^2$. Note that
in the short wave length limit we recover the standard for
plane waves: $E(k,\tau)=\hbar k$.
Also note that the energy is defined in terms of conformal
quantities (derivatives are taken with regards to $\tau$ and conformal ${\bf x}$). This means that the temperature we use must also be in the same
conformal frame: $T=T_{phys}a$ where $T_{phys}$ is the physical temperature.
If we assume that thermalization is maintained through a heat bath which
evolves as radiation, and there is no generation of entropy during
inflation, we have that $T={\rm constant}$.  This value of $T$
will be crucial in what follows.

We have not yet arrived at the final result. Any mechanism that
keeps the scalar field in thermal equilibrium must break down as a
given mode becomes larger than any causal scale, i.e. around the
horizon scale.
 On superhorizon scales, we expect the spectrum to be
frozen in---the heat bath or interactions are irrelevant. In other
words $n(k)$ will be frozen at the value it has when
$k|\tau|\simeq 1$. This means that the above expression is not
entirely accurate and we must replace $E(k,\tau)$ with
$E(k,\tau)_{|k\tau|\simeq 1}$. We can now express our main result:
the full spectrum of fluctuations, including the thermal
contribution is
\begin{equation}
{\cal P}^{Total}_{\cal R}(k)={\cal P}^{Th}_{\cal R}(k)+{\cal P}^{0}_{\cal R}(k)
={\cal P}^{0}_{\cal R}(k)[2(n(k,\tau)_{|k\tau|\simeq 1}+1] \label{thermalps}
\end{equation}
This result is slightly different from that in~\cite{rob} (the
model is not the same). It also cannot be directly compared with
the results of warm inflation~\cite{warm}.

Let us now explore the consequences of Equation \ref{thermalps}.
As a first guess,  one would expect to be
in the Rayleigh-Jeans regime when the
modes exit the horizon. We then have, on superhorizon scales,
\begin{eqnarray}
{\cal P}^{Th}_{\cal
R}(k)\simeq\frac{(p-1)^2}{4p^2-3p+1}\frac{(-\tau)^{-2\nu+1}}{z^2}\frac{k^{\frac{1+p}{1-p}}}
{\beta\pi^2}\ \label{thermal1}
\end{eqnarray}
If we reexpress Eq. \ref{thermal1}
in terms of the equation of state, we have that
\begin{eqnarray}
n_s-1=\frac{5+3w}{1+3w} \nonumber
\end{eqnarray}
Close to de-Sitter we find that $n_s\simeq 0$, that is white
noise: we do not get scale-invariance because the temperature is
decreasing like $1/a$, breaking the deSitter invariance (this is
to be contrasted with the work of~\cite{alexander}). Instead we
find that a scale invariant spectrum arises if we assume a
``phantom'' regime with $w=-5/3$.

Our expression is insensitive to the details of thermalization and horizon crossing and it gives us a reasonable idea of what to expect. A useful exercise is
to compare the contribution of thermal fluctuations relative to vacuum
fluctuations during an inflationary period. We have that
\begin{eqnarray}
\frac{{\cal P}_{\cal R}^{Th}}{{\cal P}_{\cal R}^{0}}\simeq \frac{(p-1)^2}{4p^2-3p+1}
\frac{8}{\pi |H_\nu(1)|^2}\frac{K_BT}{\hbar k} \nonumber
\end{eqnarray}
For $p\gg1$ we find
\begin{eqnarray}
\frac{{\cal P}_{\cal R}^{Th}}{{\cal P}_{\cal R}^{0}}\simeq 0.1\frac{K_BT}{\hbar k}
\nonumber
\end{eqnarray}
In general the prediction of this model is a break in the power
spectrum at pivot scale $k_p\simeq 0.1 T$. For $k<k_p$ the
fluctuations are predominantly thermal with spectral index
$n^{Th}_{s}=n^0_s-1$, with the quantum fluctuations spectral index
$n^0_s$ given by the usual formula. In this regime we are
invariably in the Rayleigh Jeans limit. For $k>k_p$ the
fluctuations are predominantly quantum, with thermal fluctuations
suppressed by a factor of $e^{-k/k_p}$, given that we are in the
Wien regime. At horizon crossing we always have $E\sim k$, so this
can be replaced in the formula for $n(k)$ in either regime.

We now examine the implications of this result for two viable
scenarios, where the fluctuations are predominantly quantum and
thermal, respectively. If we have an inflationary scenario
($w\approx -1$) then the dominant fluctuations on observable
scales should be quantum, for these are near-scale-invariant. A
priori the prediction of this model  is a turn over in the
spectral index from $n_s\approx 1$ to $n_s=0$ on large scales (for
$k<k_p$). There is clearly no evidence  for {\it higher} power in
the lowest multipoles of the CMB so, at best $k_p$ could be the
current horizon scale $k_{H0}$. This is reflected on an {\it upper
bound} on the temperature during and before inflation or
alternatively on a constraint on the ratio of the temperature
before and after reheating ($T_b$ and $T_a$). Recall that  the
conformal temperature $T=T_{phys}a$ is only a constant if there is
no entropy production, so that it does suffer a jump, from $T_b$
to $T_a$ at reheating. Bearing this in mind, $k_p\sim 0.1 T_b$ but
$k_{H_0} \sim aH_0 $.  Therefore $k_p<k_{H0}$ translates into
\be\label{bound} \frac{T_b}{T_a}< 10\times\frac{\hbar k_{H_0}}{K_B
T_0} \simeq 10^{-28} \ee A marginally tighter bound can probably
be obtained through the Grischuk-Zeldovich effect: superhorizon
fluctuations with such a red spectrum will further boost the
quadropole \cite{GrisZeld}. We can convert our constraint into a
physical temperature during inflation if we assume a specific
model. For example, if the inflation ended at the GUT scale, when
the energy scale is of order $10^{17}$ GeV and at a redshift of
$z\simeq 10^{28}$, the temperature of the Universe just before
reheating would have been, at most, $10^{-2}$ eV. This means that
the Universe hits the Planck temperature more than 68 efoldings
before reheating, so that there is scope for producing the
observed structure of the Universe (for which  $50$ to $60$
efoldings before reheating is enough), but, if the bound is
saturated, not much more. In general the bound (\ref{bound})
forces the maximum number of efoldings to be \be \label{boundef}
{\cal N}_{max}>{\cal N }_{min}+2\ln\frac{E_{Pl}}{E_{Inf}}-2.3 \ee
If we can assume that $H$ doesn't vary by much during inflation,
and if all the energy in the inflaton field is converted into
radiation during reheating we can translate the bound
(\ref{bound}) into ${\cal N}>64$. Relaxing these assumptions
produces a tighter bound. This seems to rule out open inflationary
models.

If we have a phantom scenario with $w\approx -5/3$, the observed
structure of the Universe should be thermal. The prediction is a
near scale invariant spectrum breaking into $n_s=2$ for $k>k_p$.
Thus we should have $k_p>k_{S0}$, where $k_{S0}$ is the smallest
scale for which the primordial power spectrum is observable.
The constraint is now an upper bound on how much entropy has been
produced since the observed structure left the horizon;
specifically: \be \label{boundph}\frac{T_b}{T_a}> 5
\times\frac{\hbar k_{S0}}{K_B T_0} \simeq 10^{-22} \ee where we
have assumed that the smallest scales that can be probed are of
the order of a Kpc. In this scenario we have roughly that
$a\propto 1/(-t)$, with $t<0$ (i.e. $p=-1$), $\rho\propto a^2$,
and $H^{-1}= -t$ (the horizon's physical size decreases). Also
$a=-1/(2H\tau)$.

The normalization in this model is obtained from a constant of
motion combining the energy in the thermal bath and that in the
background field. The relevant factor in (\ref{thermal1}) is
$1/(z^2\tau)$ which can be rewritten into $T^{phys}H/M_{Pl}^2$,
i.e. $\rho_{\delta\phi}^{1/4}\rho_\phi^{1/2}$, or $\sim
T^{phys}/|t|$. It's this important constant that must be $\sim
10^{-10}$ to match observations. Should all the energy in the
``phantom'' field be converted into radiation at the end of this
phase we therefore get the rather undemanding bound $T_a<10^4
M_{Pl}$ (in combination with (\ref{boundph})). But by requiring
that the current Hubble volume was once inside the phantom Hubble
volume (in a calculation mimicking the inflationary counterpart)
we find that $T^{phys}/ T_{Pl}\sim 0.1$ at the start of the
phantom phase (and that requires saturating bound
(\ref{boundph})). If the thermal bath is set up at $T^{phys}\sim
T_{Pl}$ the break into $n_s=2$ should happen only an order of
magnitude or so above $k_{S0}$. Whether this could be observed is
debatable.

Note that for simplicity we have considered $w=-5/3$, but strict
scale-invariance in this scenario is actually pathological as it
requires it is only for $-5/3<w<-1/3$ that the Newtonian potential
$\Phi$ stays constant and has the same spectrum as ${\cal R}$ on
large scales. For $w\le -5/3$ the potential diverges. However as
long as the spectrum is slightly red this is not a problem and we
have for the growing mode: \be {\cal
R}=-\frac{5+3w}{3(1+w)}\Phi\ee

We conclude with a few comments on aspects of this model, and how
they relate to other work. We stress that our system is very
different from a single thermal fluid, as previously
studied~\cite{holo,param,milne}. Here the unperturbed field
$\phi_0$ is {\it not} thermalized; only its fluctuations
$\delta\phi$ are thermalized. The fluid $\phi_0$ drives the
expansion and provides the leading order energy, but no entropy.
Whatever the equation of state $w$ for $\phi_0$, the $\delta\phi$
behave like standard thermal radiation, with $w_1=1/3$ and supply
the entirety of the entropy of the system. This feature allows us
to bypass a number of thermodynamical constraints pertaining to
single thermal fluids, namely the relation $\zeta=1+1/w$ between
the $\zeta$ exponent appearing in $\rho\propto T^\zeta$ and $w$.
If we insist on a Stephan-Boltzman law of the form $\rho_0\propto
T^\zeta$ (where $\rho_0$ is the energy in $\phi_0$, and $T$ is the
temperature of $\delta\phi$) we find instead that $\zeta=3(1+w)$.
This doesn't contradict any fundamental thermodynamical
constraint: the usual result merely indicates that the second
order energy, contained in $\delta\phi$, should go like $T^4$.

But even a two-fluid model breaks down when discussing thermal
fluctuations. Indeed Maxwell's formula, $\sigma^2_E(R)=T^2 dU/dT$,
which is the workhorse of much previous
work~\cite{hag,holo,param}, is not applicable here. The energy
fluctuation is of the form $\delta\rho\sim
\dot\phi_0\delta\dot\phi$, i.e. a cross term between the
unthermalized $\phi_0$ and the thermalized $\delta\phi$. So the
energy fluctuation of the system is, to leading order,
$\sigma^2_E(R)\propto U_0 U_1(R)$, where $U_0=\rho_0V$ is the
average energy in $\phi_0$, and $U_1(R)$ is the average energy in
$\delta\phi$ smoothed on scale $R$ (which is $\sim T$). Unusually,
we only need to know the average energy of the thermalized system
to work out the leading order energy fluctuation in the overall
system. These novelties conjure to bypass the general prediction
$n_s=4$, allowing for scale-invariant thermal fluctuations without
appealing to any new physics.

Regarding the Gaussianity of these fluctuations it has been
shown~\cite{pogo} that for a single thermal fluid thermal
fluctuations are very approximately Gaussian in the Rayleigh-Jeans
limit (but not in the Wien limit). However, just as it happens
with the equivalent calculation of the variance, the calculation
of the cumulants in a single thermal fluid is not applicable to
our system. Instead we note that the derivation of Gaussianity
usually used for linear inflation applies to any density matrix
that is diagonal in the number operator, including a thermal
state. We therefore expect the thermal component to be Gaussian,
too, rendering the thermal scenario presented above viable. This
is in contrast with non-linear inflationary couplings, that may
produce a certain degree of non-Gaussianity~\cite{mald}.

Finally, we remind the reader that we are considering a universe
that starts off in thermal equilibrium. The hallowed example is
that of what has become known as new Inflation: as the Universe
cools down, the scalar field settles down into a slow roll regime
and it is potential energy dominated. This is not, however, a
generic feature of the inflationary cosmology. One appealing
alternative is a Universe that emerges through quantum tunnelling
into an inflationary era \cite{HartleHawking}. Another possibility
is that our local patch has entered into an inflationary regime as
a result of a Planck scale fluctuation of the Inflaton
\cite{linde}. The initial state for the onset inflation would not
necessarily be thermal. In both of these scenarios we don't expect
a thermal imprint on space time on large scales.

We thank Robert Brandenberger, Chris Gordon, Kate Land and Anse
Slozar for interesting discussions. Research at PI is supported in
part by the Government of Canada through NSERC and by the Province
of Ontario through MEDT. Research at Oxford was done under the
auspices of the Beecroft Institute for Particle Astrophysics and
Cosmology.

\end{document}